\documentstyle[11pt,epsf]{article}

\def\p{\partial}
\def\f{\frac}
\def\le{\left}
\def\ri{\right}
\def\be{\begin{equation}}
\def\ee{\end{equation}}
\def\bea{\begin{eqnarray}}
\def\eea{\end{eqnarray}}
\def\nn{\nonumber}

\def\la{\lambda}
\def\t{\tau}
\def\s{\sigma}
\def\om{\omega}
\def\ve{\varepsilon}

\begin{document}

\begin{titlepage}

\vspace*{1cm}
\begin{center}
{\Large\bf Casimir energy of the Nambu-Goto string with Gauss-Bonnet
term and point--like masses at the ends$^{*}$}

\vskip 1cm
Leszek Hadasz $^{\dag}$

\vskip .5cm
Jagellonian University, Institute of Physics,\\
Reymonta 4, 30--069 Cracow,
Poland
\end{center}

\vskip 1cm

\begin{abstract}
We calculate the Casimir energy of the 
rotating Nambu-Goto string with the Gauss-Bonnet
term in the action and point-like masses at the ends. This
energy turns
out to be negative for every values of the parameters of the model.

\end{abstract}

\vspace*{\fill}

\noindent
\begin{tabular}{l}
TPJU 23/98 \\
September 1998\\
Accepted for publication in Acta Phys.Pol. {\bf B}.
\end{tabular}
\vskip 1cm

\noindent
\underline{\hspace*{8cm}}

\noindent
$^*$Work supported by the KBN grant 2 P03B 095 13. 

\noindent
$ ^{\dag}$E--mail address:  hadasz@thrisc.if.uj.edu.pl
\end{titlepage}

It seems to exist a common belief that the construction of (even approximate)
string representation of QCD could be crucial for understanding  
non--perturbative properties of quantum chromodynamics, such as 
the nature of the ground state or mechanism of confinement.
The conjecture of existence of such a description 
is supported by a number of facts \cite{intr6,polyn}, to mention only the nature
of the $1/N_c$ expansion \cite{intr1}, success of the dual models in
description of Regge phenomenology, area confinement law found
in the strong coupling lattice expansion \cite{intr3} or the existence
of flux--line solutions in confining gauge theories \cite{intr4,nkap3} and
the analytical results concerning two--dimensional QCD \cite{intr5}.
The results obtained recently in the framework of M theory are also
very promising (see, for instance, \cite{witnew}).

It is well known that the simplest,
Nambu--Goto string model \cite{intr7},
when treated as a quantum system, has many
drawbacks \cite{bnest,witt} which include 
the non--physical dimension of the space--time
(D=26) or tachion and unwanted massless states in the
spectrum. It is therefore reasonable to study modifications of the Nambu--Goto
model, and among them the simplest ones, which preserve the equations 
of motion for the interior of the string while changing the boundary
conditions imposed at the string ends.

The model we investigate in this letter is defined 
through the action functional 
\be
\label{r1}
S  =  -\int_{\t_1}^{\t_2}\!d\t\int_0^\pi\! d\s\;
\sqrt{-g}\le(\gamma +\f{\alpha}{2}R\ri)  
-  \sum_{i=1}^{2}m_i\int_{\t_1}^{\t_2}\!d\t\sqrt{\le(\p_\t X\ri)^2}.
\ee
Here $\gamma,$ with the dimension (mass)$^2,$ is the string tension, 
$\alpha$ is a dimensionless parameter and $g = {\rm det}g_{ab}$ is the 
determinant of the induced metric tensor $g_{ab} = \p_a X^\mu\p_b X_\mu$
($a,b = \t,\s$). $R,$ the inner curvature scalar, can be written in the form
$$
R= \le(g^{ab}g^{cd} - g^{ad}g^{bc}\ri)\nabla_a\nabla_b X_\mu
\nabla_c\nabla_c X^\mu,
$$
where $\nabla_a$ is a covariant (with respect to the induced metric $g_{ab}$)
derivative. Partial analysis of classical solutions of the model 
specified by (\ref{r1}) was performed in \cite{lehtom}.

The inclusion of the Gauss--Bonnet term 
$$
S_{GB} =\int_{\t_1}^{\t_2}\!d\t\int_0^\pi\! d\s\;
\sqrt{-g}\,R
$$
into the action (\ref{r1}) is a
rather natural construction in the context of the effective QCD string.
The QCD string action should contain -- apart from the $X^\mu$ fields
-- also infinitely many fields describing for instance the transverse
shape of the chromoelectric flux joining the color sources. In constructing
the effective string action, one integrates over such a fields and this
procedure inevitably leads to emergence of the intrinsic curvature term
in the action functional. Of course, it is then only the first one
out of the infinitely many terms with the growing number of derivatives.

The worldsheet parametrization can be completely fixed by imposing 
the manifestly Lorentz invariant conditions \cite{bnest}:
\bea
\label{r2}
(\dot X \pm X')^2 & = & 0, \\
(\ddot X \pm \dot X{}')^2 & = & -\f14q^2,
\eea
where the dot and the prime mean differentiation with respect to $\t$ and
$\s$ and $q$ is a parameter with the dimension
of mass. The appearance of this parameter can be traced back to the
assumption, that $\s$ takes values in the fixed interval $[0,\pi].$

It can be shown (for details see \cite{bnest,paw,paw2}), that in this
parametrization every solution of the string
equations of motion and boundary conditions, following from the action
(\ref{r1}), corresponds to the
solution of the complex Liouville equation \cite{liouv}:
\be
\label{liouv}
\ddot\Phi - \Phi'' = 2q^2e^{\Phi},
\ee
supplemented with the boundary conditions:
\be
 \label{bc}
 \le\{
 \begin{array}{lcll}
 \gamma - \alpha q^2e^{2\Re\Phi} & = &
    (-1)^im_i\f{\p}{\p\s}\le(e^{\Re\Phi/2}\ri), &  \\
 &&&  \\
 \alpha\f{\p}{\p\t}\Re\Phi & = & 0, &   \\
 &&& \hspace*{1cm} {\rm for} \hskip 5mm \s = 0,\pi. \\
 \alpha\cos\le(\Im\Phi/2\ri) & = & 0, &  \\
 &&& \\
 \f{\p}{\p\s}\Im\Phi & = & 0, &
 \end{array}
 \ri.  
\ee
 The correspondence is explicitly established 
through the relations:
\bea
\label{con1}
 e^{\Phi} & = & -\f{1}{q^2}\f{F'_L(\t+\s)F'_R(\t-\s)}{
     \sin^2\le[\f{F_L(\t+\s)-F_R(\t-\s)}{2}\ri]},  \\
 && \nn \\
\label{con2}
 X^\mu(\t,\s) & = &  X^\mu_L(\t+\s) + X^\mu_R(\t-\s),  \\
 && \nn \\
\label{con3}
 \f{\p}{\p\t}X^\mu_{L,R} & = &
 \f{q}{2|F'_{L,R}|}\le(\cosh\Im F_{L,R},\;\cos\Re F_{L,R},\;
 \sin\Re F_{L,R},\; \sinh\Im F_{L,R}\ri),
\eea
where $F_{L,R}$ are arbitrary complex functions which give single valued
$\Phi$ satisfying the boundary conditions (\ref{bc}).

A distinguished class of solutions of the Liouville equation
(\ref{liouv}) is composed of static, i.e. $\t$--independent
fields. They are of the form
\be
\label{sol1}
e^{\Phi_0} = -\f{\la^2}{q^2}\f{1}{\cos^2\le(\la\s-d\ri)},
\ee
where $\la$ and $d$ satisfy the set of algebraic equations,
\bea
\f{\gamma q}{\la^2}\cos^4d - m_1\sin d\cos^2d - \f{\alpha\la^2}{q} 
& = & 0, \nn \\
&& \\
\f{\gamma q}{\la^2}\cos^4(\pi\la -d) - m_1\sin(\pi\la-d)\cos^2(\pi\la- d)
  -  \f{\alpha\la^2}{q} & = & 0, \nn 
\eea
following from the boundary conditions (\ref{bc}) for the Liouville field 
of the form (\ref{sol1}).

The Liouville field $\Phi_0$ describes a straight string which rotates with
a constant angular velocity in some plane and by choosing a convenient 
reference frame we can write the string coordinates in a form
\be
X^\mu = \f{q}{\la^2}\Big(\la\t,\cos\la\t\,\sin(\la\s-d),
        \sin\la\t\,\sin(\la\s-d),0\Big).
\ee

Let us note, that in the presence of the inner curvature term in the 
action (\ref{r1}) the velocities of the string ends,
\bea
\label{velocity}
v_1 & = & \le|\f{dX^i}{dX^0}\ri|_{\s=0}= |\sin d|, \nn \\
v_2 & = & \le|\f{dX^i}{dX^0}\ri|_{\s=\pi}= |\sin (\pi\la -d)|,
\eea
remain smaller than the the velocity of light even in the limit of vanishing
masses $m_i=0.$

For fixed values of the external parameters $\gamma, \alpha$ and
$m_i-$s this is in fact a family of solutions, parameterized by the 
value of $q.$ By increasing $q$ we increase the
string length,
\be
L=\f{2q}{\la^2}\le[\sin^2\f{d}{2} + \sin^2\le(\f{\pi\la-d}{2}\ri)\ri],
\ee
as well as its classical energy,
\be
\label{ene}
E_0  =  \f{\pi q\gamma}{\la}\le[1+\f{\sin\pi\la\,\cos(\pi\la-2d)}{\pi\la}\ri]
        + m_1\cos d  +  m_2\cos(\pi\la-d).
\ee

In order to calculate the Casimir energy of the rotating string we have
to find the frequencies of small oscillations around this configuration.
If we write
\be
\Phi(\t,\s) = \Phi_0(\t,\s) + \Phi_1(\t,\s),
\ee
where $\Phi_0$ is given by (\ref{sol1}) and $\Phi_1$ is assumed to be small,
then from (\ref{liouv}) we get the equation
\be
\label{osc}
\p^2_\t\Phi_1 - \p^2_\s\Phi_1 + \f{2\la^2}{\cos^2(\la\s-d)}\Phi_1=0,
\ee
and (\ref{bc}) leads to the boundary conditions for the $\Phi_1$ field 
of the form
\be
\label{bc2}
\Phi_1 = 0, \hskip 5mm \Im\,\p_\s\Phi_1 = 0 \hskip 5mm {\rm for}
\hskip 5mm \s = 0, \pi.
\ee

General solution of the equation (\ref{osc}) satisfying the conditions 
(\ref{bc2}) is
\be
\label{oscsol}
\Phi_1(\t,\s)  =  \sum_{n=1}^{\infty}a_n\cos\le(\om_n\t+\phi_n\ri)
 \le[
\f{\p}{\p\s} + \la \tan(\la\s - d)\ri]\cos(\om_n\s - \delta_n),
\ee
where 
$$
\tan \delta_n = \f{\la}{\om_n}\tan d,
$$
$\om_n$ are positive roots of the equation
\be
D(\om)   \equiv    \om^2\sin\pi\om -\la\om\le[\tan d + \tan(\pi\la-d)\ri]
\cos\pi\om - \la^2\tan d\tan(\pi\la-d)\sin\pi\om  =  0,
\ee
excluding $\om_0=\la$ and $a_n, \phi_n$ are arbitrary, real constants.

It is convenient to introduce the abbreviations
$$
\eta = \la\tan(\pi\la -d), \hskip 5mm \rho = \la\tan d,
$$
what allows to rewrite
$$
D(\om) = \le(\om^2- \rho\eta\ri)\sin\pi\om -(\rho+\eta)\om\cos\pi\om.
$$

Using Eqs. (\ref{con1}--\ref{con3}) one checks that the Liouville field
$\Phi_1$ described a set of decoupled string oscillations with frequencies
\be
\label{strfre}
\nu_n=\f{\la}{q}\ \om_n.
\ee

The Casimir energy is defined as a (appropriately regularised and
renormalized) sum 
\be
\label{casdef}
E_{\rm Cas} = \le(\sum_{n=1}^{\infty}\f12\nu_n\ri)_{\rm ren}.
\ee
We choose to work with the $\zeta$
function regularization  (let us stress, however, that the
final result is independent of the chosen regularization method -- for
instance, the cut-off regularization gives the same ultimate formulae)
and define after \cite{wiwib}
\be
\label{casen1}
\tilde E_{\rm Cas} \stackrel{\rm def}{=} 
\f14\lim_{\ve\to 0}\le[\mu^{\ve}\zeta(-1+\ve) + \mu^{-\ve}\zeta(-1+-\ve)\ri]
\ee
where, for $\Re\, s > 1,$
\be
\label{zetadef}
\zeta(s) = \sum_{n=1}^{\infty}\nu_n^{-s}
\ee
and the parameter $\mu$ with dimension of mass is introduced to
ensure that the r.h.s. of the expression (\ref{casen1}) has the
dimension of energy for arbitrary complex $s.$ The physically
interesting value $s=-1$ is obtained from (\ref{zetadef}) through
the analytic continuation.

Using the standard methods of contour integration in the complex plane
one writes
\be
\label{r6}
\sum_{n}\;\nu_n^{-s} = 
\f{1}{2\pi i}\le(\f{\la}{q}\ri)^{-s}\int_{{\cal C}_1}\!dz z^{-s}
\f{d}{dz}\log D(z),
\ee
where the integration contour ${\cal C}_1$ (Fig. 1) surrounds 
zeroes of the function $D$ excluding $\nu_0 = \f{\la^2}{q}.$

\vskip 2mm

\centerline{\epsfbox{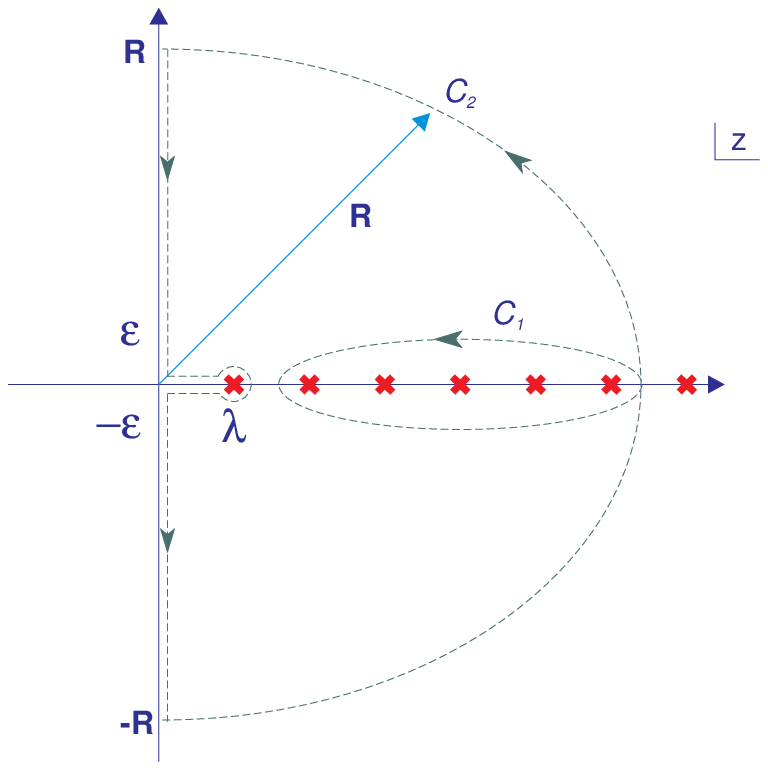}}

\vskip 2mm

\centerline{\small Fig 1. The integration contours in the complex plane.}

\vskip 2mm

The analicity of the function $D(z)$ allows to deform the integration 
contour ${\cal C}_1$ into  ${\cal C}_2$ and, after a straightforward 
calculation, one arrives at the formula
\be
\label{cef1}
\tilde E_{\rm Cas}  =  \f{\la}{2\pi q}\le[\eta^2\log\f{\eta^2}{\tilde\mu^2} + 
\rho^2\log\f{\rho^2}{\tilde\mu^2}\ri] + 
\f{\la}{2\pi q}\le\{\int_0^{\infty}\!dy\;\log\le[1-
\f{(y-\rho)(y-\eta)}{(y+\rho)(y+\eta)}{\rm e}^{-2\pi y}\ri] - \la\ri\},
\ee
where $\tilde\mu$ is also an arbitrary, but now dimensionless constant.

Following \cite{nnest1,nnest2} we interpret terms in the first square 
bracket in Eq. (\ref{cef1}) as renormalising the classical string
mass. This is also supported by the expectation, that the Casimir energy
should vanish for infinitely long strings, while the discussed terms 
fail to satisfy this condition. 

Our final expression for the Casimir energy thus reads
\be
\label{casen}
E_{\rm Cas} =
\f{\la}{2\pi q}\le\{\int_0^{\infty}\!dy\;\log\le[1-
\f{(y-\rho)(y-\eta)}{(y+\rho)(y+\eta)}{\rm e}^{-2\pi y}\ri] - \la\ri\}.
\ee

\vskip 2mm

\centerline{\epsfbox{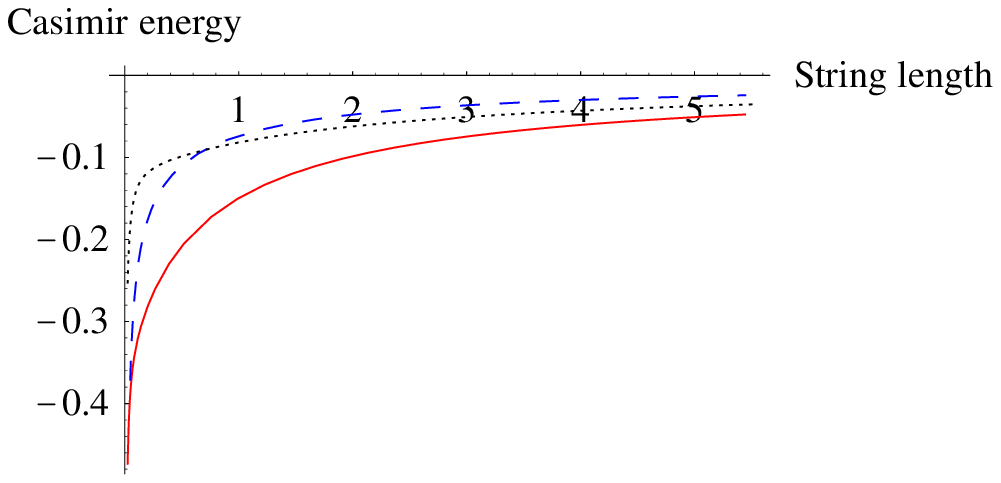}}

\vskip 2mm

\noindent
{\small
Fig 2. The Casimir energy versus string length for various values of masses 
and the parameter $\alpha:$ $\alpha = 0.2, m_1=0.1, 
m_2 = 0.2$ (solid line), $\alpha = 0.2, m_1=0.1, 
m_2 = 30$ (dashed line) and $\alpha = 2, m_1=0.1, 
m_2 = 0.2$ (dotted line). All dimensionful quantities in the system of units 
$\gamma=1.$}

\vskip 2mm
\noindent
For every values of masses $m_1, m_2$ and the parameters $\gamma, \alpha$
the Casimir energy (\ref{casen}) is negative.

For long strings $(\sqrt{\gamma}L\to\infty)$ formula (\ref{casen}) gives
$$
E_{\rm Cas} = -\f{1}{12}\f{1}{L} + o\le(L^{-1}\ri).
$$
This is different from the celebrated L\"usher term \cite{lusher},
$$
E_{C}^{\rm L} = -\f{\pi}{12}\f{1}{L},
$$
but the reasons are obvious. First, L\"usher term is derived for the
string with fixed ends and the oscillation frequencies equal
$$
\nu_{n}^{\rm L} = \f{\pi n}{L},
$$
while in our, rotating string case we have
$$
\nu_n(\sqrt{\gamma}L\to\infty) = \f{2n}{L}.
$$
Second, in considered model we have only planar oscillations
and this gives additional factor 1/2. 

\vskip 3mm

\noindent
{\Large\bf Acknowledgments.}

\noindent
This work is supported by the KBN grant 2 P03B 095 13.

\end{document}